\documentclass[a4paper,conference]{IEEEtran}

\usepackage{amsfonts}
\usepackage{amssymb}
\usepackage{graphicx} 
\usepackage{epstopdf}
\usepackage{subfigure}
\usepackage{tikz}

\graphicspath{{./figures/}}
\DeclareGraphicsExtensions{.eps}

\ifCLASSINFOpdf
\else
\fi

\usepackage[cmex10]{amsmath}
\usepackage{array}
\usepackage{mdwmath}
\usepackage{mdwtab}
\usepackage{eqparbox}
\usepackage{fixltx2e}
\usepackage{stfloats}
\usepackage{url}
\usepackage[squaren, Gray, cdot]{SIunits}

\newcommand{\SNR}{\mathsf{SNR}}
\newcommand{\Q}{\mathbb{Q}}
\newcommand{\C}{\mathbb{C}}
\newcommand{\Z}{\mathbb{Z}}
\newcommand{\F}{\mathbb{F}}
\newcommand{\K}{\mathbb{K}}
\newcommand{\argmin}{\operatornamewithlimits{argmin}}
\newcommand{\E}{\mathrm{E}}
\DeclareMathOperator{\Tr}{Tr}

\begin{document}
\title{Structured Compute-and-Forward with Phase Precoding Scheme: Beyond Integers}
\author{\IEEEauthorblockN{Ehsan E. Khaleghi}
\IEEEauthorblockA{
Communications \& Electronics Department\\
TELECOM ParisTech\\
Paris, France\\
E-mail: ehsan.ebrahimi-khaleghi@telecom-paristech.fr}
\and
\IEEEauthorblockN{Jean-Claude Belfiore}
\IEEEauthorblockA{Communications \& Electronics Department\\
TELECOM ParisTech\\
Paris, France\\
E-mail: jean-claude.belfiore@telecom-paristech.fr}}

\maketitle
\begin{abstract}
In this work, we focus on the $K-$user Gaussian Symmetric Complex-valued Interference Channels 
(GS-CIC). The Compute-and-Forward (CoF) protocol in wireless networks have been employed for 
Gaussian Symmetric Real-valued Interference Channels (GS-RIC) to achieve approximately 
the capacity of such channels and making Interference Alignment (IA) by solving a system of linear 
equations at destinations. We aim to adapt this protocol to GS-CIC. At high Signal-to-Noise 
Ratio ($\SNR$), this protocol shows a fading behavior of the achievable sum-rate for strong 
and very strong interference regimes. In complex field this random-like behavior is related 
to the phase of complex cross channel coefficients. To avoid this fading-like behavior, 
we consider $\Z[i]-$lattice codes and we propose a phase precoding scheme for CoF protocol 
with partial feedback. In this method the phase of channel coefficients will then be fed back 
to the transmitters in order to choose the best precoder factors to control this random behavior. 
We further simulate the achievable sum-rate given by this scheme and showing significant gain 
improvement can be obtained in terms of limiting the random behavior of achievable sum-rate.
\end{abstract}


\IEEEpeerreviewmaketitle
\section{Introduction}
Today the connection between different communication systems is realized via
wireless communication mediums. The most important benefit of using a wireless
medium is to have more reliable communication for users in terms of mobility, 
but it will have its own disadvantages such as communication coverage, 
interference between users (or channels), fading behavior of transmission 
channels and transmission power. Managing the interference is an issue to
overcome when users in wireless networks want to have at the same time an 
acceptable Quality of Service (QoS) and higher transmission rates.
Rather than avoiding the interference, it can be used in an intelligent way to 
recover the desired information for an intended receiver. This interference 
management technique is called IA. CoF protocol in 
wireless communication is a novel relaying strategy in which relays can decode
or compute the functions of transmitted signals from different transmitters, then
forward them to the intended destination. By using this protocol interference
could be managed by solving a system of linear equations at destinations and 
the achievable sum-rate will be higher. It can be said IA is possible via 
CoF protocol.
 
\subsection{Related Work}
Interference and its influence in communication channels have been introduced many years ago 
in \cite {IEEE:Ashlswede},\cite {Berkeley:Shannon}. Different alignment approaches 
are known in literature to manage interference in GS-RIC. Two general categories of IA are 
known as linear IA \cite{Book:Syed} and non-linear IA \cite{IEEE:Motahari}. Making IA in an 
intelligent way is one of the most important challenges in the domain of multi-user 
wireless information theory. The well known pioneers of the domain have been focused 
on the case of 2-user (or 3-user) systems to understand the importance of alignment 
approaches and its difficulties before generalizing them for the general $K-$user 
systems when $K>2$. This will led us to understand how much gain we can benefit by 
using different alignment strategies and coding methods. For $2-$user case much 
significant progress had been done for strong \cite{IEEE:Carlerial} and very strong 
interference \cite{IEEE:Jafarian} channels. In \cite{IWCIT:Ehsan}, first we have showed 
our interest to use CoF protocol as an IA technique, then we have studied deeply the 
main framework of Nazer and Gastpar scheme defined in \cite{IEEE:Gastpar}. 
We also analyzed the new method of CoF described in \cite{IEEE:Erez} 
to find the cause of random like-behavior of final achievable sum-rate showed for $2-$user 
GS-RIC. This fading behavior is limited by using Golden ratio, its equivalents and our 
proposed precoding scheme. The new method of using lattice codes over Eisentsein 
integers has been defined for CoF by Tunali \emph{et al.} \cite{IEEE:Narayan} to 
show higher information rates than reported in \cite{IEEE:Gastpar} can be 
achieved. Many research studies and code designs have been done for the case of 
real-valued interference channels, but for the complex-valued interference channels
and its code structures still we need to realize much more research. 

\subsection{Summary of Paper Results}
\begin{figure}[!h]
 \centering
 \includegraphics[width=3in]{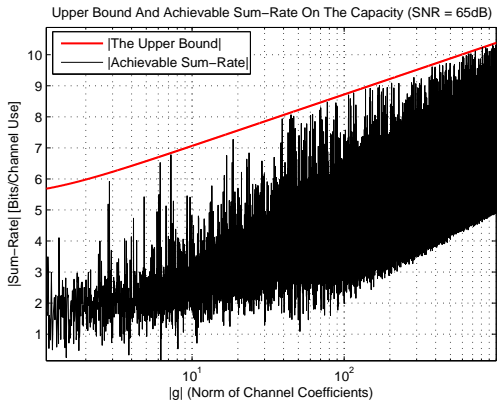}
 \caption{Upper and lower bounds on the capacity of a 2-user GS-CIC 
 with respect to the cross-gain $g=\rho e^{i\phi}$. $1 < \rho\leq10^3$ 
 and $\phi$ is chosen randomly in $\left[-\pi/4,+\pi/4\right]$ for each vaiation of $\rho$.}
 \label{fig:Fig_A}
 \end{figure}

In this paper our principal goal is to apply the computation rate defined in 
\cite{IEEE:Gastpar} by Nazer and Gastpar for GS-CIC. After defining this 
computation rate, we will use the new CoF protocol described in \cite{IEEE:Erez} 
by Ordentlich \textit{et al.} for transforming approximately the $K-$user GS-CIC to 
the $2-$user case. This transformation will help us to calculate the achievable 
sum-rate at destinations. Without using any Channel State Information (CSI) at 
transmitters the performance showed in Fig.~\ref{fig:Fig_A} is reachable by 
using the CoF protocol described in \cite{IEEE:Erez}. For the high values of 
$\SNR$, we are interested in reducing the gap between the upper bound and the 
achievable sum-rate in strong and very strong interference regimes. 
We can notice, the small variation of the interfering channel gain could 
result deep fadings on the achievable sum-rate of our transmission channel 
and it can dramatically reduce the final sum-rate.


In second part of this work we introduce channel model, lattice structures
and the complex case is defined for $2-$ user Gaussian symmetric interference 
channels (GS-IFC). In third part, the minimum mean square error (MMSE) estimator, 
the correspondent Hermitian form and the achievable sum-rate are defined by 
using the main framework of \cite{IEEE:Gastpar} and \cite{IEEE:Erez}. 
After showing the benefit of using a phase precoding scheme at transmitters, 
the new protocol of CoF combined with phase precoders is introduced. 
The phase precoders will use CSI as a feedback information to calculate 
the best phase precoder coefficients to improve the final achievable sum-rate. 
Our proposed precoding scheme also belongs to the general framework introduced 
by A. Sakzad \textit{et al.} in \cite{IEEE:Sakzad}; however, this scheme has 
not been analyzed in detail. Finally, at the last part of this work, the 
numerical results are presented for high $\SNR$ and strong interference channels. 
\subsection{Notational Convention}

Through this paper, we use $\C$, $\Z$ and $\Z\left[i\right]$ to denote the 
field of complex numbers, the set of integers and Gaussian integers, respectively. 
$\F_q$ to denote the finite field of size $q$. Here in all equations, bold letters 
are for vectors and bold capital letters are for matrices, e.g., \textbf{x} and 
\textbf{X}, respectively. We define $\log_2^{+}\left(x\right)\triangleq\max\left
(\log_2\left(x\right),0\right)$. We refer to lattices over complex integers as 
$\Z\left[i\right]-$lattices. We let $\left|z\right|$ and $\arg(z)$ denote the
modulus and the phase of the complex number $z$, respectively.

\section{channel model, lattice structure and The complex case}
 
\subsection{Channel Model}
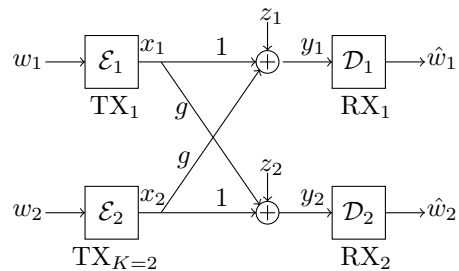
\begin{figure}[h]
 \centering
 \begin{tikzpicture}
      \draw (0,0) rectangle (0.7,-0.7);
      \node at (-0.75,-0.35) {$w_1$};
      \draw [->] (-0.52,-0.35)--(0,-0.35);  
      \node at (0.35,-0.35) {$\mathcal{E}_1$};
      \draw [->] (0.7,-0.35)--(2.25,-0.35); 
      \node at (0.4,-0.95) {$\mathrm{TX}_{1}$}; 
      \node at (0.9,-0.15) {${x}_1$};
      \draw (2.4,-0.35) circle [radius=0.15];
      \draw  (2.3,-0.35)--(2.5,-0.35);
      \draw (2.4,-0.25)--(2.4,-0.45);
      \node at (1.8,-0.15) {$1$};
      \node at (2.45,0.25) {${z}_1$}; 
      \draw [->] (2.4,0.18)--(2.4,-0.2); 
      \draw [->] (2.6,-0.35)--(3.25,-0.35); 
      \node at (3,-0.15) {${y}_1$};
      \draw (3.25,0) rectangle (3.95,-0.7);
      \node at (3.6,-0.35) {$\mathcal{D}_1$};
      \node at (3.7,-0.95) {$\mathrm{RX}_1$}; 
      \draw [->] (3.95,-0.35)--(4.47,-0.35); 
      \node at (4.7,-0.3) {$\hat{w}_1$};
                              
      \draw (0,-2) rectangle (0.7,-2.7);
      \node at (-0.75,-2.35) {${w}_2$};
      \draw [->] (-0.52,-2.35)--(0,-2.35);  
      \node at (0.35,-2.35) {$\mathcal{E}_2$};
      \draw [->] (0.7,-2.35)--(2.25,-2.35); 
      \node at (0.4,-2.95) {$\mathrm{TX}_{K=2}$}; 
      \node at (0.9,-2.15) {${x}_2$};
      \draw (2.4,-2.35) circle [radius=0.15];
      \draw  (2.3,-2.35)--(2.5,-2.35);
      \draw (2.4,-2.25)--(2.4,-2.45);
      \node at (1.8,-2.15) {$1$};
      \node at (2.45,-1.75) {${z}_2$}; 
      \draw [->] (2.4,-1.82)--(2.4,-2.2); 
      \draw [->] (2.55,-2.35)--(3.25,-2.35); 
      \node at (3,-2.15) {${y}_2$};
      \draw (3.25,-2) rectangle (3.95,-2.7);
      \node at (3.6,-2.35) {$\mathcal{D}_2$};
      \node at (3.7,-2.95) {$\mathrm{RX}_2$}; 
      \draw [->] (3.95,-2.35)--(4.47,-2.35); 
      \node at (4.7,-2.3) {$\hat{w}_2$};           
                         
      \draw [->] (1,-0.35)--(2.3,-2.25); 
      \draw [->] (1,-2.35)--(2.3,-0.45); 
      \node at (1.27,-1) {$g$}; 
      \node at (1.3,-1.65) {$g$};
 \end{tikzpicture}
 \caption{2-user Gaussian Symmetric complex-valued Interference Channel.}\label{Fig:GS-CFC}
 \end{figure}

The $K-$user GS-CIC is the general model of this paper. The transmitters
reliably communicate linear functions to $M$ relays over a complex-valued 
symmetric interference channel. In this model each relay $m$ observes a noisy 
linear combination of the transmitted signals through the channel,

\begin{equation}
 \label{Channel_Mod}
 {y_m}=\sum \limits_{l=1}^K h_{ml}\cdot{}{x}_{l}+{z}_{m}
\end{equation}
where $h_{ml}\in\C$ are the channel coefficients, 
$\mathbf{x}=\left[x_{1},\cdots,x_{K}\right]$, $\mathbf{y}=\left[y_{1},\cdots,y_{K}\right]$ 
and $\mathbf{z}=\left[z_{1},\cdots,z_{K}\right]$ are used to denote the 
input-vector, the output-vector and the complex i.i.d Gaussian 
noise-vector all of size \emph{K}, respectively. Furthermore, 
${\mathbf{z}_m}\sim\mathcal{CN}\left(0,{\sigma}^2\right)$ and 
$\mathbf{x}_{l}\in\Lambda_{c}$. Let $\mathit{\Lambda_{c}}$ 
be the coding lattice which is common to all users.
In this paper all users have the same power constraint i.e., 
$P_\mathrm{i} = P$, so the $\SNR$ is defined as 
$\SNR=\frac{P}{\sigma^2}$. By considering 
a simple lattice IA defined in \cite{IEEE:Erez} this $K-$user 
case will be approximately equivalent to a $2-$user case showed
in Fig.~\ref{Fig:GS-CFC}, this means that,

\begin{equation}
\mathbf{H}=\left[\begin{array}{cc}
1 & g\\
g & 1
\end{array}\right]
\end{equation} 
which the channel coefficients are considered to be in the field of complex 
numbers.

The simplified expression of (\ref{Channel_Mod}) for $2-$user case is written as,
\begin{equation}
 \label{Sim_Chan_mod}
 \left[\begin{array}{c}y_1\\y_2\end{array}\right]=
 \mathbf{H}\cdot\left[\begin{array}{c}x_1\\x_2\end{array}\right]
 +\left[\begin{array}{c}z_1\\z_2\end{array}\right]
\end{equation}

What we see in Fig.~\ref{Fig:GS-CFC}, the channel is symmetric, it means 
$\mathbf{H}(i,j) = g$ for all $i \not= j$, and $\mathbf{H}(i,i) = 1$ for all \emph{i}'s.

In addition, each transmitter must satisfy the power constraint, 
which for \emph{n} channel uses and user \emph{i} is given by:

\begin{equation}
\label{Chan_power_cont}
{\frac{1}{n}} \sum_{j=0}^{n} \vert {\mathbf{x}_\mathrm{\emph{i}}}\vert^{2} \leq P_\mathrm{\emph{i}}
\end{equation}

In order to perform CoF, we need a couple of ingredients:

\begin{itemize}
 \item \emph{Each} transmitter is equipped with an encoder $\mathcal{E}_{l}$, that maps length$-K$ messages 
 over a \emph{lattice code}.
 \item \emph{Each} receiver will recover the transmitted messages by using a decoder $\mathcal{D}_{l}$,
 that maps the received messages over a \emph{lattice code} to its transmitted version. This mapping is 
 possible with \emph{K} integers $\{a_{1},\cdots,a_{K}\}$ which will \textit{approximate}
 the channel gains $h_{ml}$. The main idea is that, if $\mathbf{x}_{l}\in\Lambda_{c}$ and if $\Lambda_{c}$
 is a coding complex lattice, then
 \begin{equation}
 \lambda\triangleq\sum \limits_{l=1}^K a_{l}\mathbf{x}_{l}\in\Lambda_{c}
 \end{equation}
 as well.
\end{itemize}

\subsection{Lattice structure}

We recall some fundamental notation of lattice codes which are essential throughout the 
paper. The definition of a \emph{N-dimension complex lattice $\Lambda$} with 
\emph{generator matrix} $\mathbf(G)\triangleq\left(\mathbf{g}_{1}^T|
\mathbf{g}_{2}^T|\dots|\mathbf{g}_{N}^T\right)$, for all $\mathbf{g}_{j}\in\C^{m}$ and
$1\le j\le N$ is the set of points in $\C^{m}$ (the $\mathbf{g}_{i}^T$ vectors 
are considered to be the columns of $\mathbf{G}$)

\begin{equation}
\label{Lattice_Model}
\Lambda =  \{\mathbf{x}=\mathbf{u}\mathbf{G}|\mathbf{u}\in{\Z\left[i\right]^{m}}\}
\end{equation}
when $m=N$, the lattice is called \emph{full rank}. Around each lattice point 
$\mathbf{x}\in\Lambda$ there is a \emph{Voronoi region} 
$\nu\left(\mathbf{x}\right)$ defined as:

\begin{equation}
\label{Ver_region}
\nu\left(\mathbf{x}\right) = \{\mathbf{y}\in\C^{m}:\|\mathbf{y}-\mathbf{x}\|\le
\|\mathbf{y}-\lambda\|, \forall\lambda\in\Lambda\}
\end{equation}

If $\Lambda'$ is a lattice itself, then a subset $\Lambda'\subseteq\Lambda$ is called 
a \emph{sublattice}. By having a sublattice $\Lambda'$, the \emph{lattice code} $\Lambda/\Lambda'$
can be define. $\Lambda/\Lambda'$ includes a finite constellation of lattice points from the 
lattice $\Lambda$. An intelligent choice \cite{IEEE:Sloane} for the sublattice $\Lambda'$ is $a\Lambda$ for some 
$a\in\Z\left[i\right]$. The shape of the lattice constellation is determined by the Voronoi region
of the lattice $\Lambda'$. we can also define the \emph{lattice modulo operation} as

\begin{eqnarray*}
{Q_\Lambda} {\left(\mathbf{x}\right)}\triangleq\argmin_{\lambda\in\Lambda} \|\mathbf{y}-\mathbf{\lambda}\| \\
\mathbf{y}\mod\Lambda\triangleq\mathbf{y}-{Q_\Lambda} {\left(\mathbf{x}\right)}
\end{eqnarray*}

The generator matrix of a lattice can be found by applying the Cholesky decomposition to a definite positive 
Hermitian form of a complex lattice.

\subsection{The Complex Case}

In the complex case, we have another degree of freedom. Now, instead of dealing with $\Z-$lattices, 
we consider an $\mathcal{O}_{\K}$-lattice $\Lambda_{c}$ where $\mathcal{O}_{\K}$ is the ring of 
integers of a complex quadratic number field $\K=\Q(\sqrt{-b})$ for some square free positive 
integers $b$. When $\mathcal{O}_{\K}$ is a Principal Ideal Domain (PID), then the definition 
of an $\mathcal{O}_{\K}$-lattice is similar to the one of a $\Z-$lattice, and there exists 
lattice basis. Unfortunately, when $\K$ is a complex quadratic number field, then 
$\mathcal{O}_{\K}$ is not in general a PID. It is a PID when $b\in\{1,2,3,7,11,19,43,67,167\}$. 
The most natural choice, in the complex case , is to mimic the real case, that is take $b=1$, 
which implies $\mathcal{O}_{\K}=\Z[i]$. A more judicious choice has been made in \cite{IEEE:Narayan}
with $b=3$ giving rise to $\mathcal{O}_{\K}=\Z[\omega]$ where $\omega=\zeta_{3}$ 
\footnote{Here, as usually, $\zeta_{q}$ denotes a primitive $q^{th}-$root of unity.}. 
As a lattice, $\Z[\omega]\simeq A_{2}$, that is the hexagonal lattice, when $\Z[i]\simeq\Z^{2}$. 
This is explained by the fact that $A_{2}$ is better for quantization than $\Z_{2}$.

\section{MMSE estimator and The Compute-and-Forward With Phase precoding}

Suppose that we code over \emph{n} time-slots where \emph{n} is the degree of the relative 
number field $\K$ over quadratic complex field $\F$ which can be, for instance, $\Z[i]$ or 
$\Z[\omega]$. In this part first we aim at computing the MMSE, $n\times n$ matrix B, 
Then we define our proposed phase precoding scheme to improve 
the behavior of the achievable sum-rate.   

\subsection{the minimum mean square error estimator}

We suppose, for sake of simplicity, that all transmitted signals $\mathbf{x}_{k}$ are coming 
from a 1-dimensional $\mathcal{O}_{\K}-$lattice, which is obviously $\mathcal{O}_{\K}$ itself. 
Let $Gal_{\K/\F}=\{\sigma_0,\sigma_1\,\dots,\sigma_{n-1}\}$ denotes the Galois group of the 
extension $\K$ over $\F$ and set

\begin{equation}
\mathbf{A}_{k}=diag(\sigma_0(a_k),\sigma_1(a_k),\dots,\sigma_{n-1}(a_k))
\end{equation}

Where $a_k\in\mathcal{O}_{\K}$. In the same way, the transmitted vectors are 
$\mathbf{x}_{k}=diag(\sigma_0(x_k),\sigma_1(x_k),\dots,\sigma_{n-1}(x_k))$, with 
$x_k\in\mathcal{O}_{\K}$. The received vector is expressed in (\ref{Channel_Mod}).
Now we multiply (\ref{Channel_Mod}) by the MMSE matrix B and get,

\begin{equation}
\label{MMSE_mult}
\mathbf{B}\mathbf{y}=\sum \limits_{k=1}^K h_{k}\mathbf{B}\mathbf{x}_{k}+\mathbf{B}\mathbf{z}
\end{equation}

But we want to decode the equation $\sum\limits_{k=1}^K a_{k}\mathbf{x}_{k}$, which means, 
the lattice point $\sum\limits_{k=1}^K h_{k}\mathbf{B}\mathbf{x}_{k}$ through canonical
embedding. So, equation (\ref{MMSE_mult}) becomes,

\begin{equation}
\label{MMSE_mult_0}
\mathbf{B}\mathbf{y}=\sum\limits_{k=1}^K \mathbf{A}_{k}\mathbf{x}_{k}+\tilde{\mathbf{z}}
\end{equation}
where $\tilde{\mathbf{z}}=\sum\limits_{k=1}^K (h_{k}\mathbf{B}-\mathbf{A}_{k})
\mathbf{x}_{k}+\mathbf{B}\mathbf{z}$ is the equivalent noise. Minimizing the mean square 
error is equivalent to minimize,

\begin{equation}
\begin{array}{c}
\label{MMSE_mult_1}
\varepsilon=\frac{1}{n}\E\left[\tilde{\mathbf{z}}^{\dagger}\tilde{\mathbf{z}}\right]\\
=\frac{P_{x}}{n}\sum\limits_{k=1}^K \Tr\left[(h_{k}\mathbf{B}-\mathbf{A}_{k})(h_{k}
\mathbf{B}-\mathbf{A}_{k})^{\dagger}\right]+N_{0}\Tr\left[BB^{\dagger}\right]
\end{array}
\end{equation}

where $P_{x}$ is the per user power and $N_0$ is the per component noise variance. 
Let us now rewrite the equation (\ref{MMSE_mult_1}), 

\begin{equation}
\begin{array}{c}
\varepsilon=\frac{P_{x}}{n}\alpha \Tr\left[\mathbf{B}\mathbf{B}^\dagger-
\mathbf{B}\mathbf{C}^{\dagger}-\mathbf{C}\mathbf{B}^{\dagger}+\frac{1}{\alpha}
\sum\limits_{k=1}^K \mathbf{A}_{k}\mathbf{A}_{k}^\dagger\right]\\
=\varepsilon=\frac{P_{x}}{n}\alpha \Tr\left[(\mathbf{B}-\mathbf{C})
(\mathbf{B}-\mathbf{C})^{\dagger}+\frac{1}{\alpha}
\sum\limits_{k=1}^K \mathbf{A}_{k}\mathbf{A}_{k}^{\dagger}-
\mathbf{C}\mathbf{C}^{\dagger}\right]
\end{array}
\end{equation}
with $\alpha=\|h\|^{2}+\frac{1}{\rho}, \rho=\frac{P_{x}}{N_{0}}$ and
\begin{equation}
C=\frac{1}{\alpha}\sum\limits_{k=1}^K h_{k}^{*}\mathbf{A}_{k}
\end{equation}

The minimization of $\varepsilon$ with respect to $\mathbf{B}$ is equivalent to set 
$\mathbf{B}=\mathbf{C}$, in which case the expression of $\varepsilon$ is

\begin{equation}
\begin{array}{c}
\label{MMSE_mult_2}
\varepsilon=\frac{P_{x}}{n} \Tr\left[\sum\limits_{k=1}^K \mathbf{A}_{k}\mathbf{A}_{k}^{\dagger}
-\frac{1}{\alpha}\left(\sum\limits_{k=1}^K h_{k}^{*}\mathbf{A}_{k}\right)
\left(\sum\limits_{k=1}^K h_{k}\mathbf{A}_{k}^\dagger\right)\right]\\
=\frac{p_{x}}{n} \{\sum\limits_{k=1}^K\Tr_{\K/\F}\left(|a_{k}|^{2}\right)
-\frac{1}{\alpha}\sum\limits_{i,j=1}^K h_{i}h_{j}^{*}\Tr_{\K/\F}\left(a_{i}a_{j}^{*}\right)\}.
\end{array}
\end{equation}

\subsection{The Compute-and-Forward With Phase Precoding}

In terms of making IA, we aim to not divide the general GS-CIC to two separate 
channels in parallel (create from real and imaginary parts of the original 
complex-valued channels). The objective is to keep the GS-CIC in its original 
shape, this choice will intercept of having interference between parallel 
channels. The same CoF protocol introduced in \cite{IEEE:Erez} is used in 
this section for GS-CIC. We propose a phase precoding scheme which is using 
channel coefficients as feedback to improve the results of previous works. 
In CoF protocol the computation rate is known as a maximal rate which users can
send their codewords to a destination, this rate is given by:

\begin{equation}
\label{Computation_rate}
R(\mathbf{h},\mathbf{a})=\frac{1}{2}\log_2^{+}\left\{\left({\parallel \mathbf{a}
\parallel^{2}}-\frac{\mathsf{SNR}|\mathbf{h}^{*}\mathbf{a}|^{2}}{1+\mathsf{SNR}
{\parallel \mathbf{h}\parallel^{2}}}\right)^{-1}\right\}
\end{equation}
where $\mathbf{a}$ is a vector of Gaussian integers characterizing the equation we want to decode. 

We are interested to improve the behavior of achievable sum-rate given by CoF protocol
in the \emph{strong} and \emph{very strong} interference regimes. For $2-$user 
GS-CFC, $\mathbf{h}=[1,g]$ and the interference-to-noise ratio is defined as 
$\mathsf{INR}\triangleq |g|^{2}\mathsf{SNR}$. Here, $g=\rho e^{i\phi}\in\C$ is
the cross channel coefficient and $\mathbf{a}=[a_1,a_2]$ with $a_1,a_2\in\Z[i]$. 
The bloc fading channel model is considered, then the simplified expression of 
(\ref{Computation_rate}) is

\begin{equation}
\label{Computation_rate_form}
R(\mathbf{h},\mathbf{a})=\frac{1}{2}\log_{2}^{+}\left\{ \frac{\left(\frac{1}
{\mathsf{SNR}}+(1+|g|^{2})\right)}{q\left(a_1,a_2\right)}\right\} 
\end{equation}

where for a complex-valued channel $q(a_1,a_2)$ is a definite positive Hermitian form equal to:

\begin{equation}
\label{quadratic_form}
\begin{array}{c}
q\left(a_1,a_2\right)=(|a_1g|-|a_2|)^{2}+\frac{1}{\mathsf{SNR}}(|a_1|^{2}+|a_2|^{2})\\
-2\Re({a_{1}a_{2}^{*}g})+2|a_{1}a_{2}g|
\end{array}
\end{equation}

We can deduce the Hermitian matrix of (\ref{quadratic_form}) as, 

\[G= \left(\begin{array}{ c c } |g|^{2}+\frac{1}{\mathsf{SNR}} & -g \\ 
-g^* & 1+\frac{1}{\mathsf{SNR}} \end{array} \right)\]

\begin{figure}[!h]
 \centering
 \includegraphics[width=3in]{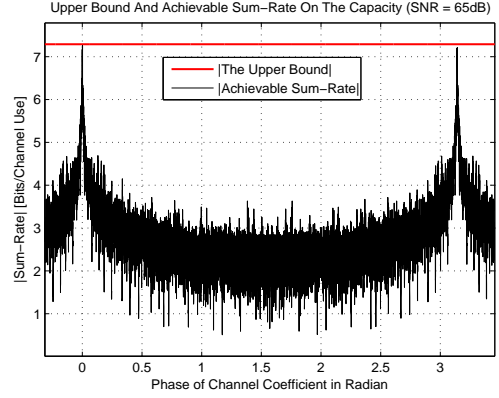}
 \caption{Upper and lower bounds on the capacity of a 2-user GS-CIC 
 with respect to the phase variation $\phi$. $g=\rho e^{i\phi}$ with $0\le\phi\le\pi$ 
 and $\rho=7.0770$ is set to be unchangeable.}
 \label{fig:Fig_B}
 \end{figure}

As a definite positive Hermitian form, $q(a_1,a_2)$ defines a rank $2$ complex-lattice,
$\Lambda_{\mathrm{CF}}$. we aim at finding the two successive minima of (\ref{quadratic_form}).
The Hermitian form is of dimension $2$, an algorithm for finding the two
successive minima is the Complex-LLL algorithm defined by \cite{IEEE:Cong}. This algorithm will reduce 
the basis of a lattice in complex field. Let define matrix $B=\mathbf{Cholesky}(G)$ be a basis of 
$\Lambda_{\mathrm{CF}}$ and $B_{\mathrm{red}}$ be the reduced basis after complex-LLL reduction. 
$U$ is the Unimodular basis change matrix. Call $G_{\mathrm{red}}=B_{\mathrm{red}}^T B_{\mathrm{red}}$ 
the reduced Gram matrix, then the two successive minima are the diagonal entries of $G_{\mathrm{red}}$. 
If we fix the $\rho$ part of the cross channel coefficient, and varying its phase between 
$[0,\pi]$, it can be observed in Fig.~\ref{fig:Fig_B} when the phase is approximately around 
$0$ or $\pi$ the achievable sum-rate will be maximal. In this scenario $\arg(g)$ 
should be aligned to reach the maximal achievable sum-rate. 

After analyzing the method described in \cite{IEEE:Sakzad}, we 
propose our precoding scheme. In our scheme each codeword should be multiplied by a 
precoder before transmitting them to destinations, this will let us to align their phase to maximize
the final achievable sum-rate. To realize this scheme we assume that the precoder for each 
transmitter is a complex scalar $e^{i\theta_k}$, for some $-\pi/4\le\theta_k\le+\pi/4$, multiplying 
the lattice codeword. Finding the optimal precoder factors is realized by using the CSI available as
feedback information at transmitters. We apply the phase precoding function as,
\begin{equation}
\label{Phase_precoder_1}
\begin{array}{c}
 D_{k}:\C^n\to\C^n\\
 D_{k}\left(\mathbf{x}_k,\theta_{k}\right)\triangleq e^{i\theta_{k}}\mathbf{x}_k
\end{array}
\end{equation}

for $\theta_{k}\in\left[-\pi/4,\pi/4\right]$ and $1\le k\le K$. In the $2-$user case, we send 
$e^{i\theta_1}x_{1}$ and $e^{i\theta_2}x_{2}$ instead of $x_1$ and $x_2$. The phase precoded 
$e^{i\theta_{k}}\mathbf{x}_k$ continues to satisfy the power constraint (\ref{Chan_power_cont}).
In general the equivalent channel coefficients vector is,
\begin{equation}
\label{Phase_precoder_2}
\begin{array}{c}
 \mathbf{h}'=\left[h_{1}e^{i\theta_1},h_{2}e^{i\theta_2}\right]\\
 =\left[\rho_{1}e^{i\times\left(\phi_{1}+\theta_1\right)},\rho_{2}e^{i\times\left(\phi_2+\theta_2\right)}\right]
\end{array}
\end{equation}
When the channel is symmetric, $h_{1}=1$. In our proposed precoding scheme 
after applying precoders, the new channel coefficients are $\mathbf{h}'=
\left[e^{i\left(\theta_1\right)},\rho_{2}e^{i\left(\phi_2+\theta_2\right)}\right]$
, the optimal choice for precoders is therefore,

\begin{equation}
\begin{array}{l}
 \theta_1=0\\
 \theta_2=-(\phi_{2}\pm\eta)
\end{array}
\end{equation}
$\eta\in\left[-0.2~\radian,+0.2~\radian\right]$ is the precoder precision factor, it is used to
keep $\left(\phi_2+\theta_2\right)$ approximately in an area which the rate shown 
in Fig.~\ref{fig:Fig_B} is maximal. We dispose the CSI, then calculating $\theta_2$ 
is feasible by using four quadrant $\arctan$ function, $\theta_2=\arctan_2(\Im(g)/\Re(g))$. 
The numerical result of this scheme is presented in the next section to show its performance. 

\section{Simulation Results}
The proposed scheme defined in section I\hspace{0.08mm}I\hspace{0.08mm}I was 
implemented with following parameters: The channel is considered as a 2-user 
GS-CIC with $h_1=1$ and $h_2=\rho e^{i\phi}$. To calculate the sum-rate, $\SNR$ 
is chosen to be 65 dB and the lattice reduction is done in the complex field. 
For evaluating the performance of our proposed scheme and being in strong and 
very strong interference regimes, $\rho$ is set to vary in $\left[1.1,10^3\right]$ 
with a very high precision. Then $\phi$ is selected ramdonly in 
$\left[-\pi/4,\pi/4\right]$ for each variation of $\rho$. 
The reason for choosing $\phi$ in this interval, 
is because of using $\Z\left[i\right]$-lattices. 

 \begin{figure}[!h]
 \centering
 \includegraphics[width=3in]{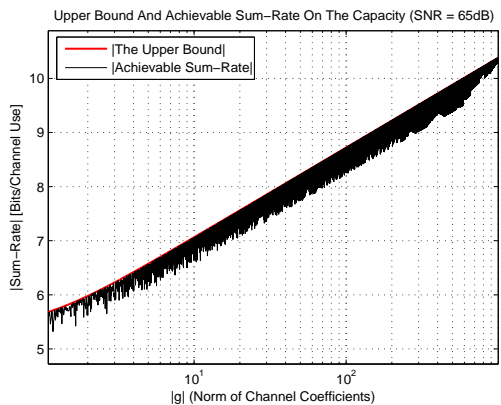}
 \caption{Upper and lower bounds on the capacity of a 2-user GS-CIC 
 with respect to the cross-gain. Our proposed Phase precoding technique is used.}
 \label{fig:Fig_C}
 \end{figure}
It can be observed from Fig.~\ref{fig:Fig_C} that for this configuration, 
the phase precoding scheme defined in previous section improve the achievable 
sum-rate of the CoF protocol \cite{IEEE:Erez} showed in Fig.~\ref{fig:Fig_A} 
for the case of 2-user GS-CIC. As we can see in Fig.~\ref{fig:Fig_C}, the deep 
fadings have been limited to 0.3 [Bits/Channel Use], and there is no more gap 
between the upper bound and the achievable sum-rate. It means by using this 
scheme the interference have been aligned and also the performance of the CoF 
protocol has been improved.

\section{Conclusion}
In this paper, after studying the channel model, $\Z\left[i\right]-$lattice 
structure and the MMSE estimator matrix for the complex-case, we have showed 
the existence of a phase precoding scheme with partial feedback for the 
CoF protocol defined in \cite{IEEE:Gastpar}, which will increase the finale 
communication rate between users. This proposed precoding scheme is enabling 
the relays to decode reliably to linear combination of lattice points with 
complex integer coefficients. Numerical results suggest that using our 
precoding scheme will limit the fading-like behavior and also achieve 
significant gain improvement in the case of 2-user GS-CIC as presented in 
\cite{IEEE:Erez}. Choosing the best precoder factor with a very high precision
is essential step to improve performance of the CoF protocol when the interference 
is available. Future work will include a more detailed study of precoding schemes 
in the case of lattices over Eisenstein integers.

\end{document}